\documentclass{article}

\usepackage{arxiv}

\usepackage[utf8]{inputenc} 
\usepackage[T1]{fontenc}    
\usepackage{hyperref}       
\usepackage{url}            
\usepackage{booktabs}       
\usepackage{amsfonts}       
\usepackage{nicefrac}       
\usepackage{microtype}      
\usepackage{lipsum}
\usepackage{graphicx}

\usepackage{tabularx}
\usepackage{multirow}

\graphicspath{ {./images/} }

\title{Bridging Modalities: Joint Synthesis and Registration Framework for Aligning Diffusion MRI with T1-Weighted Images}

\author{
 Xiaofan Wang \\
  University of Electronic Science and \\
  Technology of China,\\
  Chengdu, China\\
   \And
 Junyi Wang \\
  University of Electronic Science and \\
  Technology of China,\\
  Chengdu, China\\
  \And
 Yuqian Chen \\
  Brigham and Women’s Hospital,\\
  Harvard Medical School,\\
  Boston, USA\\
  \And
 Lauren J. O’Donnell \\
  Brigham and Women’s Hospital,\\
  Harvard Medical School,\\
  Boston, USA\\
  \And
 Fan Zhang\thanks{Corresponding author: Fan Zhang, fan.zhang@uestc.edu.cn}\\
  University of Electronic Science and \\
  Technology of China,\\
  Chengdu, China\\    
}

\begin{document}
\maketitle
\begin{abstract}
Multimodal image registration between diffusion MRI (dMRI) and T1-weighted (T1w) MRI images is a critical step for aligning diffusion-weighted imaging (DWI) data with structural anatomical space. Traditional registration methods often struggle to ensure accuracy due to the large intensity differences between diffusion data and high-resolution anatomical structures. This paper proposes an unsupervised registration framework based on a generative registration network, which transforms the original multimodal registration problem between b0 and T1w images into a unimodal registration task between a generated image and the real T1w image. This effectively reduces the complexity of cross-modal registration. The framework first employs an image synthesis model to generate images with T1w-like contrast, and then learns a deformation field from the generated image to the fixed T1w image. The registration network jointly optimizes local structural similarity and cross-modal statistical dependency to improve deformation estimation accuracy. Experiments conducted on two independent datasets demonstrate that the proposed method outperforms several state-of-the-art approaches in multimodal registration tasks. 
\end{abstract}


\section{Introduction}
Medical image registration is a key component of multimodal image fusion and plays an essential role in neuroimaging research. Accurate alignment between diffusion MRI (dMRI) and T1-weighted (T1w) images enables dRMI-derived features -- such as fractional anisotropy (FA) maps and fiber tractography -- to be spatially aligned within a standard anatomical space, facilitating population-level analyses, brain atlas construction, and clinical planning \cite{Radwan2022-pc, Hagmann2007-ib}. However, large intensity and contrast differences between modalities still pose challenges for traditional optimization-based and existing deep learning methods \cite{Wang2018-ex, Wang2024-ym}.

Early medical image registration mainly relies on intensity-based similarity measures combined with optimization strategies to estimate rigid or non-rigid transformations \cite{Klein2010-ob, Haber2007-vj}, but achieving robust and accurate alignment in multimodal settings remains challenging \cite{Haskins2020-mp, Wu2016-pv}. In recent years, unsupervised deep learning methods have gained significant attention due to their ability to learn without manual registration labels or ground-truth deformation fields \cite{De_Vos2019-qb, Sokooti2017-dr, 9665765}. A representative example is the VoxelMorph framework \cite{Balakrishnan2019-pf}, which predicts the deformation field from a moving image to a fixed image using a UNet architecture and employs a spatial transformer network (STN) \cite{Jaderberg2015-bw} for alignment. While these methods significantly improve efficiency, they still encounter performance bottlenecks in cross-modal tasks due to differences in intensity distributions \cite{CHEN2025103385}.

To address the cross-modal registration challenges, recent studies are interested in leveraging intermediate representations or latent feature spaces to improve registration accuracy \cite{Dey2022-iw, Han2022-sd}. For example, SynthMorph \cite{Hoffmann2022-bx} trains a registration network on synthetic images to construct a contrast-invariant registration framework. However, such image transformation methods may introduce artificial anatomical features and hallucinated artifacts. Other approaches that explore cross-modal correspondences in latent feature spaces \cite{CHEN2025103385}, such as the CoMIR model \cite{NEURIPS2020_d6428eec}, often require rigorous feature decoupling and involve complex network design and training processes.

Recently, joint synthesis-registration frameworks have attracted increasing attention \cite{Chen2022-qp,Kim2021-bk}. These methods integrate image synthesis and registration into a unified modeling process, aiming to produce more stable and structurally consistent deformation estimations. For instance, TransMorph \cite{Chen2022-qp} leverages a Transformer-based architecture to model long-range spatial dependencies, while CycleMorph \cite{Kim2021-bk} incorporates cycle-consistency losses for joint optimization. Nevertheless, existing methods still struggle to maintain anatomical fidelity and deformation stability, and none have directly targeted the unique challenges of dMRI-to-T1w registration.

In this work, we design a novel unsupervised joint synthesis and registration framework for dMRI-T1w registraion to bridge different modalities of medical images. The method includes a synthesis module to synthesize intermediate modality images (T1w$^{\prime}$) resembling T1w from the anatomical representations learned from dMRI, thereby transforming the original cross-modal registration task (dMRI $\rightarrow$ T1w) into a unimodal registration task (T1w$^{\prime}$ $\rightarrow$ T1w). A registration module then aligns the generated T1w$^{\prime}$ images with the real T1w images, allowing stable deformation field estimation using conventional similarity measures. The entire network is trained in an unsupervised manner, jointly optimizing image similarity and deformation smoothness. It can iteratively leverage the registration outputs as priors to guide the fine-tuning of the synthesis model,
enabling a bi-directional optimization within the synthesis-registration loop. We evaluate the method on two independent MRI datasets, and the results demonstrate that the two-stage synthesis-registration strategy consistently outperforms existing state-of-the-art methods across multiple quantitative metrics and visual assessments.

\section{Method}
\label{sec:method}

\begin{figure}[hb]
  \centering
  \centerline{\includegraphics[width=14cm]{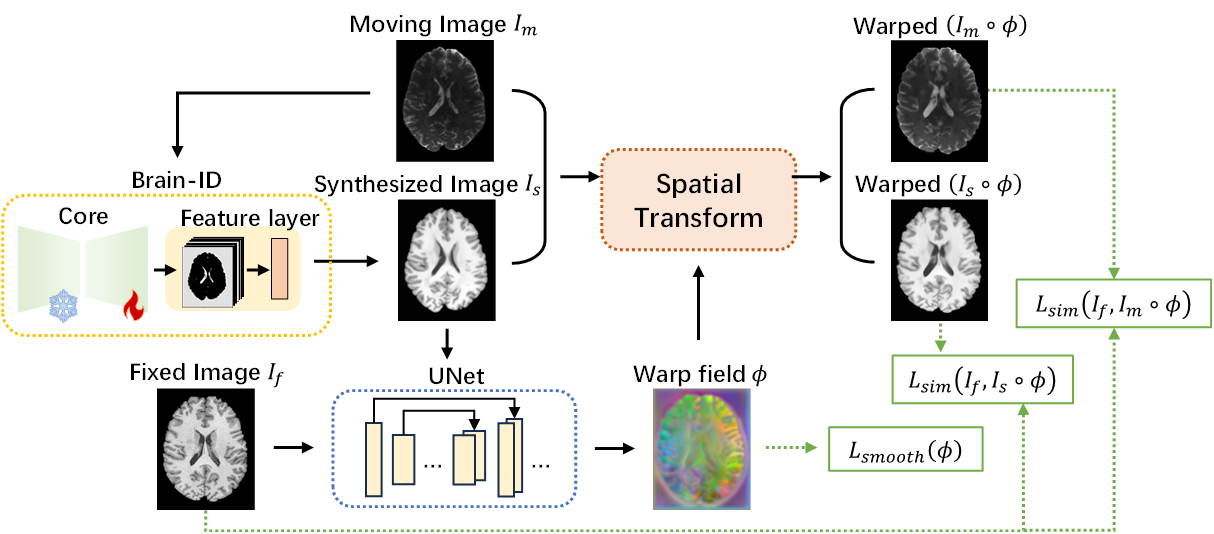}}
  \vspace{-5pt}
\caption{Framework of the proposed method.}
\label{fig:framework}
\end{figure}

The goal of our method is to compute an optimal spatial transformation $\phi$ that aligns the moving image $I_m$ (dMRI) to the fixed image $I_f$ (T1w), such that the deformed image $I_m \circ \phi$ is anatomically consistent and similar to $I_f$. Figure \ref{fig:framework} gives an overview of our method. The network takes $I_m$ and $I_f$ as inputs. Based on the pretrained Brain-ID \cite{liu2024brain} model as backbone, the synthesis module first converts $I_m$ into a structurally consistent synthetic image, $I_s$ (T1w$^{\prime}$). Subsequently, $I_s$ and $I_f$ are fed into a UNet-based registration network to predict the deformation field $\phi$. The spatial transformer module then warps the original $I_m$ into the space of $I_f$ according to $\phi$. During training, the network computes consistency losses between $I_m \circ \phi$ and warped synthetic image $I_s \circ \phi$ with respect to the fixed image $I_f$, along with regularization terms enforcing the smoothness and morphology of the deformation field. These combined constraints guide the end-to-end optimization of the framework, achieving synergistic synthesis and registration.

\subsection{Image Registration}
\label{sec:registration}

Brain-ID is an anatomical representation learning model for brain imaging, pretrained on a large-scale, multi-contrast dataset from a single subject \cite{liu2024brain}. It is capable of synthesizing brain images that preserve anatomical structures while exhibiting contrast-robust features. In our framework, the synthesis module inherits this architecture and takes the b0 image from dMRI as input to generate a synthetic T1w$^{\prime}$ image. Specifically, the input b0 volume is divided into 3D patches of size 128×128×128, each of which is encoded to extract local anatomical features and subsequently reconstructed into a complete T1w$^{\prime}$ image. The resulting image preserves the anatomical structures of b0 image while exhibiting intensity characteristics similar to those of the real T1w image.

During fine-tuning, the feature extraction backbone of Brain-ID is frozen to preserve its anatomical representation capabilities, while the decoder and task-specific head are unfrozen and optimized to enhance the discriminability of the synthetic image for the registration task. The synthesis module leverages the normalized cross-correlation (NCC) loss of T1w and warped T1w$^{\prime}$ as an indirect supervision signal, thereby achieving task-driven optimization of the generator for subsequent registration.

\subsection{Image Generation}
\label{sec:generation}

The registration module utilizes a UNet encoder \cite{Balakrishnan2019-pf} to process the synthesized image T1w$^{\prime}$ and the real T1w image as inputs. The network learns deformation features between the generated and fixed images, and outputs a 3D deformation field $\phi$, which is applied to T1w$^{\prime}$ and b0 image via a spatial transformer module network to obtain the registered images.

During training, the registration network minimizes a combination of loss functions to optimize the deformation field, including normalized cross-correlation and mutual information (MI). To learn the deformation differences between the moving and fixed images, the spatial transformer network applies the estimated field $\phi$ to warp the input.

To ensure spatial continuity and physical plausibility of the deformation, a local smoothness constraint is imposed, preventing folding or discontinuities in the transformation. As in \cite{Balakrishnan2019-pf}, the smoothness regularization $L_{{smooth}}$ is computed as the $l_2$ norm of the gradient of the final deformation field $\phi$ at each voxel $p$, with an additional $l_2$ penalty applied to the deformation field itself. The total loss is defined as follows equation:
\begin{equation}
    L_{total} = L_{sim}(I_f, I_m \circ \phi) + \lambda_1 L_{sim}(I_f, I_s \circ \phi) + \lambda_2 L_{smooth}(\phi)
    \label{eq:loss}
\end{equation}

Throughout training, the generation and registration modules are jointly updated in each batch, enabling mutual refinement. This iterative synergy allows the model to progressively generate more accurate T1w$^{\prime}$ images, ultimately improving the system’s overall performance on cross-modal registration tasks.

\section{Experiments}
\label{sec:experiments}

\subsection{Datasets}
\label{sec:datasets}

We used 180 paired dMRI and T1w brain MRI scans from two publicly available datasets: the Human Connectome Project (HCP) \cite{GLASSER2013105} and the Parkinson’s Progression Markers Initiative (PPMI) \cite{Marek2011-xk}. In the HCP, 90 scans of healthy young adults were acquired on a customized Connectome Siemens Skyra scanner (dMRI: TE = 89.5 ms, TR = 5520 ms, voxel size = 1.25 mm$^3$, 288 volumes with 18 b0 and 270 diffusion-weighted images at b = 1000/2000/3000 s/mm$^2$). Corresponding T1w images (TE = 2.14 ms, TR = 2400 ms, voxel size = 0.7 mm$^3$) were using rigid registered to MNI space via the HCP minimal preprocessing pipeline.
In the PPMI, 90 scans were used, including 65 Parkinson’s disease patients and 25 healthy controls, acquired on a 3T Siemens Trio scanner (dMRI: TE = 88 ms, TR = 7600 ms, voxel size = 2 mm$^3$, 65 volumes with 1 b0 and 64 diffusion-weighted images at b = 1000 s/mm$^2$; T1w: TE = 2.98 ms, TR = 2300 ms, voxel size = 1 mm$^3$).

For both datasets, paired dMRI and T1w scans were randomly divided into training (n = 50), validation (n = 20), and testing (n = 20) subsets. During preprocessing, images from both datasets were first resampled to an isotropic resolution of 1.25 mm$^3$. Subsequently, HCP images were cropped to 128 × 160 × 144 voxels, while PPMI images were cropped to 144 × 176 × 144 voxels.

\subsection{Implementation}
\label{sec:implementation}

The proposed framework was implemented in PyTorch (v2.0.1), employing the pretrained Brain-ID as the synthesis backbone and a UNet-based network for registration. End-to-end training was performed using the ADAM optimizer with a base learning rate of 1×10$^{-5}$. To balance optimization across modules, the decoder and task-specific head were trained with 0.1× the base rate, while the registration network used the full rate. All experiments were conducted on a workstation equipped with an NVIDIA GeForce RTX 3090 GPU and CUDA 12.2. The source code will be made publicly available upon publication.

\subsection{Experimental Design and Evaluation Metrics}
\label{sec:metrics}

Comparative experiments included traditional registration methods and representative deep learning approaches: ANTs (SyN) \cite{AVANTS200826}, two VoxelMorph variants (VXM\_NCC and VXM\_MI) trained with NCC and MI losses, and the SynthMorph model \cite{Hoffmann2022-bx}. To assess the effect of joint optimization in the proposed synthesis–registration framework, we conducted an ablation experiment. The parameters of the synthesis module were fixed so that its output remained unchanged during training, and only the registration network was optimized. The resulting frozen synthesis variant was compared with the joint framework to investigate the performance gain from joint training.

We evaluated registration performance by comparing the spatial overlap of anatomical segmentations between T1w and dMRI after registration. Specifically, T1w images were segmented using FreeSurfer \cite{Fischl2012-fo}, and dMRI were segmented using DDparcel \cite{Zhang2024-gd}. The Dice score was used to quantify volumetric overlap between corresponding anatomical structures, providing a measure of how well the segmented regions coincide after registration. In addition, the average surface distance (ASD) was employed to evaluate the boundary alignment by computing the mean distance between the surfaces of the registered and reference segmentations. While Dice reflects region-level consistency, ASD captures surface-level geometric accuracy. Higher Dice and lower ASD values indicate better registration performance.

\section{Results and Discussion}
\label{sec:discussion}

\begin{table}[t]
\centering
\footnotesize
\caption{Ablation study on the effect of the joint framework.}
\vspace{3pt} 
\begin{tabular*}{0.6\textwidth}{@{\extracolsep{\fill}} c|ccc }
\hline
Metric & VXM\_MI & Frozen-synthesis & Ours \\
\hline
Dice (\%) & 0.893 ± 0.006 & 0.912 ± 0.003 & \textbf{0.922 ± 0.004} \\
ASD (mm) & 1.150 ± 0.087 & 0.922 ± 0.065 & \textbf{0.849 ± 0.063} \\
\hline
\end{tabular*}
\label{ablation}
\end{table}

\begin{figure}[t]

  \centering
  \centerline{\includegraphics[width=14cm]{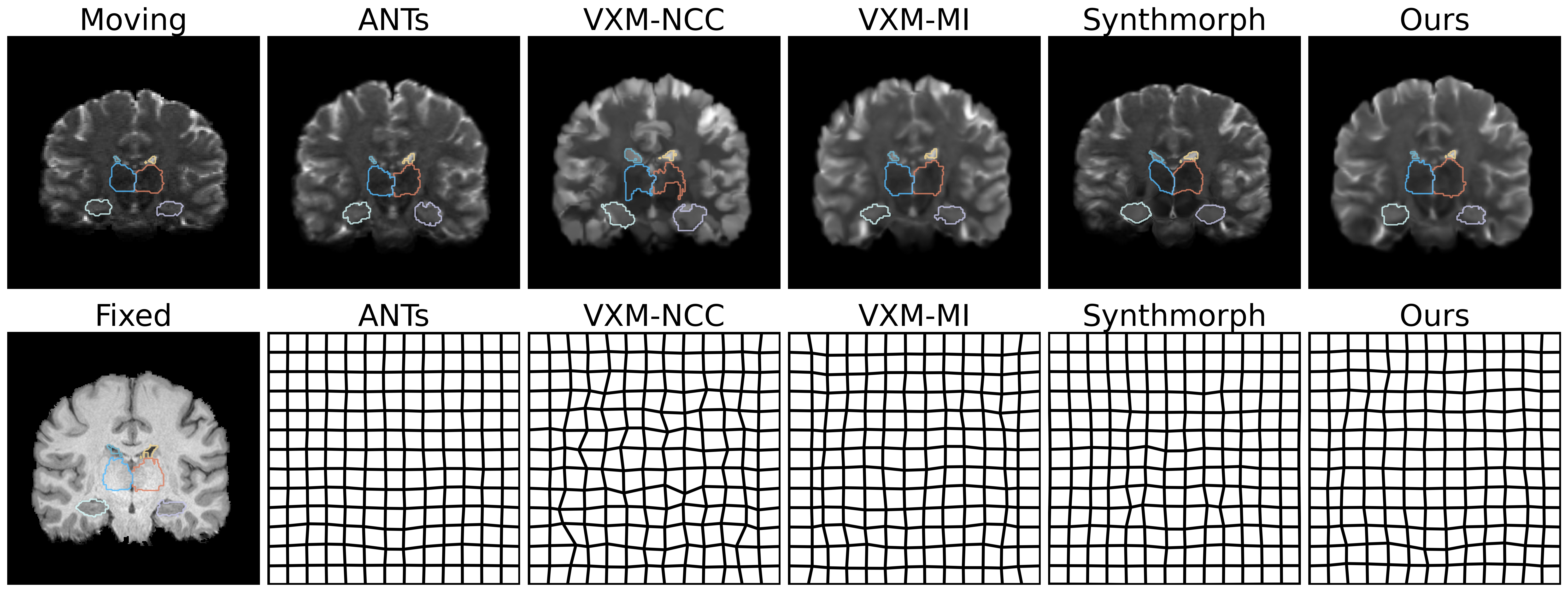}}
  \vspace{-5pt}
\caption{Warped images (row 1, columns 2–6) and the corresponding instance deformation fields $\phi$ (row 2, columns 2–6) after registering the moving images to the fixed images (column 1). Boundaries of the ventricles, thalamus, and hippocampus are overlaid to illustrate the plausibility of the deformations produced by our method.}

\label{warped imgs}
\end{figure}

An ablation study was conducted to evaluate the effectiveness of the proposed joint synthesis–registration framework, comparing the full method with the frozen-synthesis variant and VXM\_MI on HCP dataset. Dice coefficient and ASD were used as quantitative metrics to evaluate whole-brain structures. As shown in Table \ref{ablation}, the proposed framework achieved the highest Dice and lowest ASD, indicating superior performance. Experiments show that although frozen synthesis outperforms direct registration, it is still inferior to joint optimization, demonstrating that the structural prior provided by the generation module is crucial for registration accuracy.

\begin{table*}[t]
\centering
\footnotesize
\caption{Quantitative Results on the HCP and PPMI Datasets.}
\vspace{3pt} 
\begin{tabular}{>{\centering\arraybackslash}p{1.2cm}|>{\centering\arraybackslash}p{0.8cm}|>
{\centering\arraybackslash}p{2cm}|>{\centering\arraybackslash}p{1.8cm}>{\centering\arraybackslash}p{1.8cm}>{\centering\arraybackslash}p{1.8cm}>{\centering\arraybackslash}p{1.8cm}>{\centering\arraybackslash}p{1.8cm}>{\centering\arraybackslash}p{1.8cm}}
\hline
Metric & Dataset & Region & ANTs & VXM\_NCC & VXM\_MI & Synthmorph & Ours \\
\hline
\multirow{6}{*}{\shortstack{Dice (\%)}}  & \multirow{3}{*}{HCP} & cerebral WM & 0.668 ± 0.028 & 0.751 ± 0.012 & 0.706 ± 0.018 & 0.739 ± 0.031 & \textbf{0.759 ± 0.018}\\
 &  &  cerebral cortex & 0.542 ± 0.018 & 0.563 ± 0.024 & 0.605 ± 0.019 & 0.587 ± 0.038 & \textbf{0.664 ± 0.013}\\
 &  &  whole brain & 0.869 ± 0.006 & 0.853 ± 0.010 & 0.893 ± 0.006 & 0.875 ± 0.016 & \textbf{0.922 ± 0.004}\\

 & \multirow{3}{*}{PPMI} & cerebral WM & 0.779 ± 0.015	& 0.767 ± 0.011 & 0.778 ± 0.011 & 0.810 ± 0.013 & \textbf{0.815 ± 0.014}\\
 &  &  cerebral cortex & 0.638 ± 0.022	& 0.660 ± 0.020 & 0.665 ± 0.021 & 0.651 ± 0.018 &\textbf{0.674 ± 0.022}\\
 &  &  whole brain & 0.899 ± 0.010	& 0.902 ± 0.010 & 0.902 ± 0.011 & 0.900 ± 0.010 & \textbf{0.904 ± 0.012}\\
\cline{1-3}

\multirow{6}{*}{\shortstack{ASD (mm)}}  & \multirow{3}{*}{HCP} & cerebral WM & 1.201 ± 0.083	& 1.066 ± 0.107 & 1.011 ± 0.079 & 1.032 ± 0.180 & \textbf{0.856 ± 0.061}\\
 &  &  cerebral cortex & 0.863 ± 0.047	& 0.767 ± 0.070 & 0.712 ± 0.052 & 0.815 ± 0.171 & \textbf{0.576 ± 0.032}\\
 &  &  whole brain & 1.476 ± 0.096	& 1.275 ± 0.124 & 1.150 ± 0.087 & 1.309 ± 0.273 & \textbf{0.849 ± 0.063}\\

 & \multirow{3}{*}{PPMI} & cerebral WM & 0.738 ± 0.074	& 0.574 ± 0.054 & 0.550 ± 0.056 & 0.607 ± 0.075 & \textbf{0.475 ± 0.066}\\
 &  &  cerebral cortex & 0.587 ± 0.061	& 0.453 ± 0.052 & 0.447 ± 0.056 & 0.531 ± 0.070 & \textbf{0.436 ± 0.069}\\
 &  &  whole brain & 0.890 ± 0.111	& \textbf{0.650 ± 0.100} & 0.654 ± 0.103 & 0.841 ± 0.126 & 0.684 ± 0.119\\
\hline

\end{tabular}
\label{all}
\end{table*}

Figure \ref{warped imgs} visualizes the deformation fields and segmentation results of selected structures on an example HCP dataset. ANTs produces smooth deformation fields but achieves insufficient structural alignment; VXM\_NCC generates less smooth fields with poor alignment; while VXM\_MI and SynthMorph show improvements and align certain structures well (such as the hippocampus), but still exhibit limitations in complex brain regions. In contrast, our method achieves superior structural alignment and anatomical integrity while maintaining overall smoothness.

The quantitative comparison results are presented in Table \ref{all}. Across both datasets, our method consistently outperforms others in Dice and ASD, showing stable performance in whole-brain regions as well as in white matter and cortical gray matter, which are large and well-defined areas. ANTs shows the poorest performance on the HCP dataset but performs relatively better on the PPMI dataset, which can be attributed to the smaller deformations between b0 and T1w images in PPMI. Results from VoxelMorph trained with different similarity metrics suggest that it benefits from the MI loss in cross-modal registration. SynthMorph achieves stable alignment but exhibits limitations under large deformations. Notably, VXM\_NCC shows moderate Dice performance on the PPMI dataset but achieves the best ASD, likely due to residual non-brain tissue and exaggerated, irregular deformations affecting surface distance calculations. 

\begin{figure}[t]

  \centering
  \centerline{\includegraphics[width=14cm]{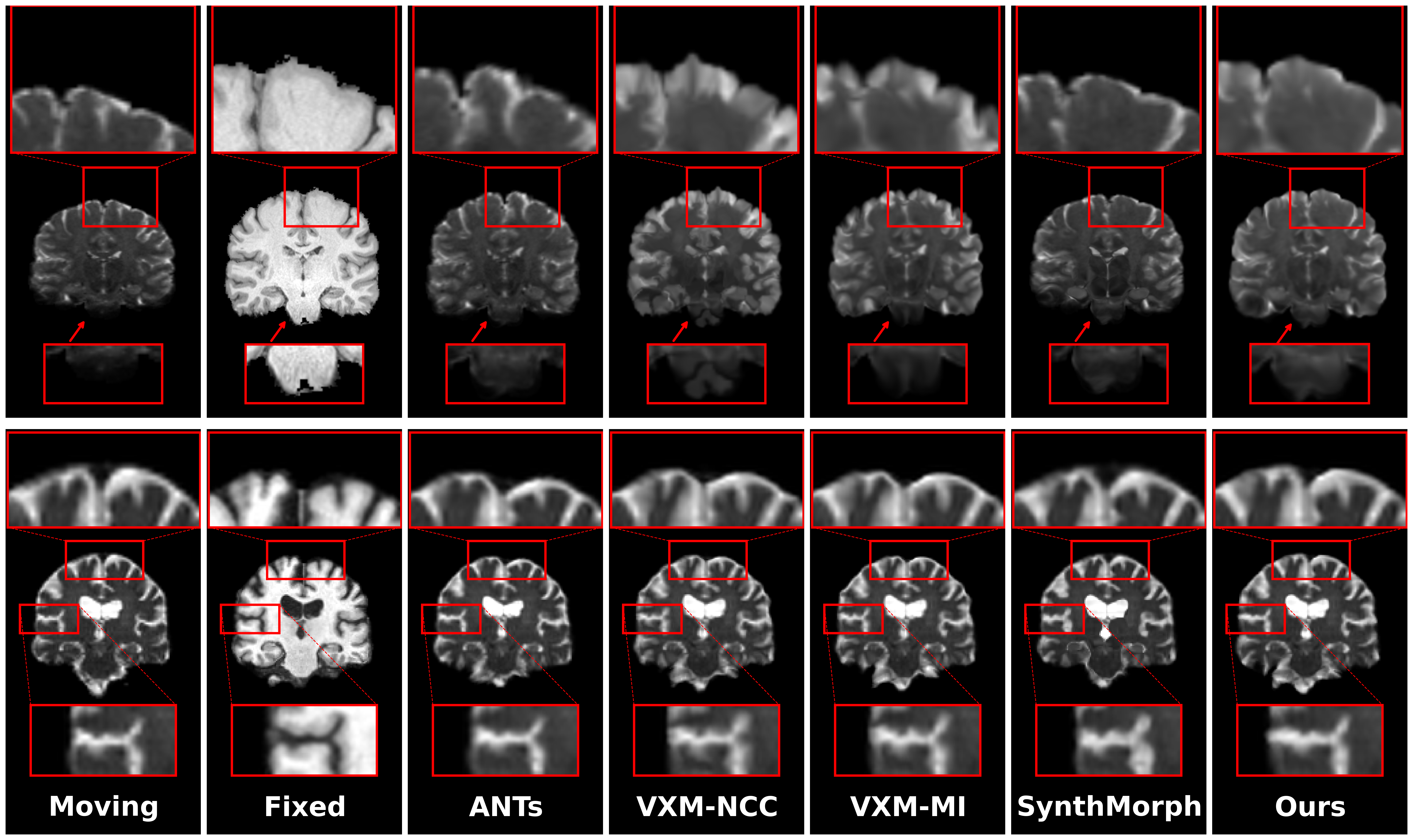}}
  \vspace{-5pt}
\caption{Visual comparison of registration methods on two datasets. Each row shows the warped images alongside the original and fixed images, with red boxes highlighting fine structural details.}

\label{zoomed imgs}
\end{figure}

Figure \ref{zoomed imgs} presents typical coronal slices from both datasets. Differences are notable in the frontal and parietal cortices, cortical folds, and brainstem, as highlighted by the red boxes. In HCP, ANTs produces slight inward shifts of anatomical boundaries and misaligns frontoparietal regions. SynthMorph preserves central brain structures under large deformations but compresses superior brain regions, resulting in locally flattened structures. This is less evident on PPMI, though inter-sulcal gaps in cortical areas are wider. VXM\_NCC introduces artifacts and anatomically implausible boundaries under large deformations, while VXM\_MI partially mitigates these issues. Both VoxelMorph variants show limited frontoparietal alignment across the two datasets. In contrast, our method consistently preserves overall brain morphology and aligns local structures, especially in the cortical folds and frontoparietal regions, consistent with the quantitative results.

Overall, our method transforms cross-modal registration into a unimodal task through a generative module, enabling the network to focus on learning geometric deformations while achieving robust performance on data with large deformations or heterogeneity. Although currently still based on pairwise registration and relying on a limited number of modalities, this framework provides a foundation for future multimodal registration and group template construction.

\section{Conclusion}
\label{sec:conclusion}

We propose a multimodal unsupervised registration framework based on a synthesis–registration collaboration. It synthesizes an anatomically consistent T1w$^{\prime}$ image to reduce structural distortion, then performs an unimodal registration with the real T1w image. Joint optimization of the synthesis and registration modules enables learning task-driven similarity metrics while avoiding excessive anatomical deformation, effectively mitigating intensity and structural inconsistencies in cross-modal registration. Experimental results on two datasets demonstrate that our approach achieves superior registration performance compared to state-of-the-art methods.

\section{Compliance with ethical standards}
\label{sec:ethics}

This research study was conducted retrospectively using human subject data made available in open access by HCP and PPMI. Ethical approval was not required as confirmed by the license attached with the open access data.

\section{Acknowledgments}
\label{sec:acknowledgments}

This work is in part supported by the National Key R\&D Program of China (No. 2023YFE0118600) and the National Natural Science Foundation of China (No. 62371107).

\bibliographystyle{unsrt}  
\bibliography{references}  

\end{document}